\documentclass[11pt,preprint]{aastex}
\usepackage{lscape}
\begin{document}

\title{NGC6362: the least massive globular cluster with chemically distinct 
multiple populations
\footnote{Based on data obtained at the Very Large Telescope under the programs 
073.D-0211 and 093.D-0618.}}

\author{Alessio Mucciarelli$^{2}$,
Emanuele Dalessandro$^{2}$,
Davide Massari$^{3,4}$,
Michele Bellazzini$^{3}$,
Francesco R. Ferraro$^{2}$,
Barbara Lanzoni$^{2}$,
Carmela Lardo$^{6}$,
Maurizio Salaris$^{6}$,
Santi Cassisi$^{5}$}

\affil{$^{2}$Dipartimento di Fisica \& Astronomia, Universit\`a 
degli Studi di Bologna, Viale Berti Pichat, 6/2 - 40127
Bologna, Italy}

\affil{$^{3}$INAF - Osservatorio Astronomico di Bologna, Via Ranzani 1 - 40127
Bologna, Italy}

\affil{$^{4}$Kapteyn Astronomical Institute, University of Groningen, Landleven 12, 9747 AD 
Groningen, The Netherlands}

\affil{$^{5}$INAF - Osservatorio Astronomico di Teramo, via Mentore Maggini, 64100, Teramo, Italy}

\affil{$^{6}$Astrophysics Research Institute, Liverpool John Moores University, 146 Brownlow Hill, 
Liverpool L3 5RF, United Kingdom}

\begin{abstract}
We present the first measure of Fe and Na abundances in NGC~6362, 
a low-mass globular cluster where first and second generation stars 
are fully spatially mixed. A total of 160 member stars (along the red giant branch and 
the red horizontal branch) have been observed with the multi-object spectrograph 
FLAMES at the Very Large Telescope. We find that the cluster has an iron abundance of [Fe/H]=--1.09$\pm$0.01 
dex, without evidence of intrinsic dispersion. On the other hand, the [Na/Fe] distribution 
turns out to be intrinsically broad and bimodal. The Na-poor and Na-rich stars populate, respectively, the bluest 
and the reddest red giant branches detected in the color-magnitude diagrams including the U filter. 
The red giant branch is composed of a mixture of first and second generation stars in a similar proportion, 
while almost all the red horizontal branch stars belong to the first cluster generation. 
Until now, NGC~6362 is the least massive globular cluster where both the photometric and spectroscopic 
signatures of multiple populations have been detected.

\end{abstract}


\section{Introduction}

Globular clusters (GCs) have long been considered the best example of a 
{\it Simple Stellar Population}, i.e., stellar systems formed of stars with the same age and initial chemical
composition (Renzini \& Buzzoni 1986). This traditional paradigm remains still valid to
a certain extent, although a wealth of recent photometric and spectroscopic results have shown that GCs are 
not as simple as previously thought, harboring multiple stellar populations (MPs) 
that differ in terms of chemical abundances.

Star-to-star differences in the C and N abundances have been known for decades in GCs 
\citep[see e.g.][]{osborn71,norris81,martell09,pancino10}, with the detection of 
CN-weak and CN-strong stars. 
In the last years the use of high-resolution spectroscopy coupled with 
large samples of GC stars has established that old, massive GCs, 
both in our Galaxy \citep{carretta09,gratton12r} and in the Local Group satellites 
\citep{m09} show star-to-star variations in some light elements: 
intrinsic scatters of Na and O abundances have been observed in all the GCs, 
variations of Al abundances in most of them, while star-to-star scatters in Mg abundances only in some 
peculiar clusters.

MPs in GCs manifest themselves also with different features in the color-magnitude
diagrams (CMDs) when appropriate bands (or filter combinations) are used: main sequence splittings, 
as those observed in $\omega$~Centauri \citep{bedin04}, NGC~2808 
\citep{piotto07} and NGC~6752 \citep{milone10}, sub-giant branch (SGB) splittings 
\citep[see e.g.][]{milone08,piotto12} and, in the majority of the cases, multimodal red giant branches 
\citep[RGBs, see][for a recent review]{piotto15}.

The most commonly accepted idea about MP formation is that 
secondary generations are formed from the ejecta of first generation stars (polluters) 
along with {\sl pristine material} 
(i.e. material with the same abundances as the first generation).  
Within this general framework, four main scenarios have been proposed differing mainly for the nature of the polluters: 
i) asymptotic giant branch stars \citep{dercole08}, ii) fast-rotating massive stars \citep{decressin07},
 iii) interacting massive binary stars \citep{demink09}, and (iv) super-massive stars \citep{denis14,denis15}.
An alternative scenario that does not require an age difference between first and second generation stars is 
the so-called {\sl early disk accretion} scenario \citep{bastian13}. This model suggests that interacting massive stars
and binaries can shed enriched material into the cluster, and low-mass pre-main sequence stars (of the same 
generation), which are fully convective and have proto-planetary disks, 
are able to sweep up the enriched material and eventually accrete it.  However, all the proposed scenarios present 
some shortcomings and they are not able to reproduce simultaneously 
all the observational pieces of evidence \citep[see e.g.][]{bastian15,bastian15b}.

By using a proper combination of archival 
Hubble Space Telescope (HST) and WFI@MPG/ESO data, \citet[][hereafter PaperI]{dalex14} found that 
in all CMDs involving the U band or the F336W HST filter, 
the RGB and SGB of the GC NGC~6362 split in two distinct sequences. 
On the contrary, in the (V, V-I) CMD the average color spread of RGB and SGB stars is fully compatible 
with the photometric errors. 
Theoretical models demonstrate \citep{sbordone11} that this behavior is the photometric signature of 
the CNONa chemical anomalies.
The RGB and the SGB split in the (U, U-V) CMD because of star-to-star variations in C and N abundances 
which are detectable 
by the U band (through the NH and CN bands present in this spectral range), 
while pure optical filters are insensitive to such variations. 
Nowadays, this behavior of the evolutionary sequences of GCs in U and optical filters is 
a common feature of GCs \citep[see e.g][]{lardo11}.

As discussed in PaperI, NGC~6362 appears to be peculiar in many respects. 
With an estimated mass of $\sim5\cdot10^{4}M_{\odot}$, 
NGC~6362 is the least massive GC with 
detected photometric MPs. 
These observational findings 
put strong constraints to the mass threshold enabling the formation of MPs. 
In addition we have found that NGC6362 is the first cluster where 
the radial distributions of its MPs show no significant radial difference up to its tidal radius.
We interpreted such an unexpected evidence as the result of a very advanced dynamical evolution and
possibly of a significant mass-loss due to interactions with the Galactic disk and potential well 
\citep[see ][]{dalex15}.

However, while the presence of the RGB photometric split indicates the presence 
of MPs in this cluster, only a dedicated spectroscopic investigation can reveal 
the specific chemical patterns associated to each sub-population needed for a
proper comparison with other systems. 
Therefore, in order to fully characterize the MPs of NGC~6362, we have collected 
high-resolution spectra for a large number of member stars. 
In this paper we present the first derivation of the Fe and Na abundances ever obtained for this cluster, 
and we discuss these results in light of those provided in PaperI.

The paper is structured as follows: in Section~2 the data-set is presented, in Section~3 we describe the details
of the adopted data-analysis procedures, in Section~4 we present the derived abundances of Na and Fe for
the stars in our sample, in Section~5 we discuss these results and in Section~6 we compare them to what 
observed in  M4, which shares a similar metallicity. 
The main results are then summarized in Section~7.

\section{Observations}
The observations have been performed with the multiplex facility 
FLAMES@ESO-VLT \citep{pasquini} in the UVES+GIRAFFE combined mode (Prop ID: 093.D-0618, PI: Dalessandro). 
This mode allows the simultaneous allocation of 8 UVES high-resolution fibers 
and 132 mid-resolution GIRAFFE-MEDUSA fibers. 
The employed gratings are the UVES Red Arm CD\#3 580, that covers the spectral 
range between $\sim$4800 and $\sim$6800 \AA\ with a spectral resolution of $\sim$45000, 
and the GIRAFFE setups HR11 (5597-5840 \AA\ and R$\sim$24000) and HR13 
(6120-6405 \AA\ and R$\sim$22000). 
All the employed setups allow to derive abundances of Fe (thanks to several 
available lines) and Na (sampling the two Na doublets at 5682-5688 \AA\ and 6154-6160 \AA\ ). 
Unfortunately, oxygen abundances cannot be derived because of the close-to-zero 
radial velocity of the cluster leading to a blending between the forbidden oxygen 
line at 6300.3 \AA\  and the sky emission line at the same wavelength.

\begin{figure*}[h]
\plotone{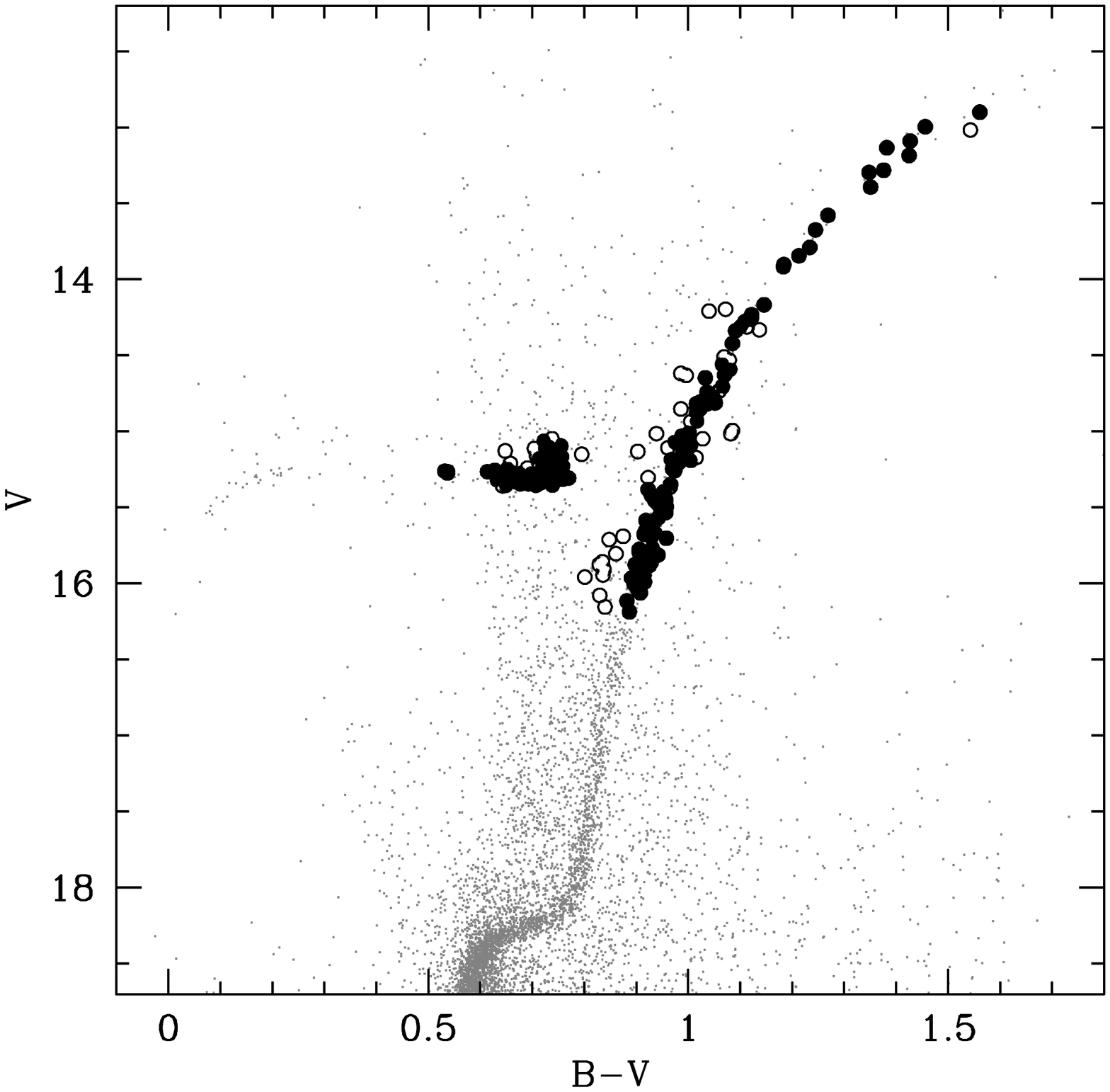}
\caption{(V, B-V) CMD of NGC~6362 (small grey circles) with marked the position of the spectroscopic targets, 
filled circles are the cluster members, while the empty circles the targets flagged as field stars. 
Photometry from \citet{dalex14}.}
\label{cmd0}
\end{figure*}

A total of 2 exposures of 45 min each for two different star configurations has been secured 
for the grating HR11 and 2 exposures of 32 min each have been performed with the grating HR13. 
The UVES targets have been kept fixed on the same targets in both star configurations in order 
to increase the S/N ratio. A total of 219 stars have been selected along the RGB and the red Horizontal 
Branch (RHB) from the ACS@HST and WFI@2.2m-ESO catalogues presented in PaperI.
These stars are
(i) brighter than V=16.2, thus reaching at least S/N ratio of about 30 per pixel;
(ii) isolated, i.e. stars without a companion brighter than V$<V_{star}+$1.0 within 2'';
and, for RGB targets only, (iii) along the two RGBs observed in the 
(U, U-V) plane, thus to properly sample the two observed sub-populations detected in PaperI. 
The position of the spectroscopic targets in the (V, B-V) CMD is shown in Fig.~\ref{cmd0}.
Table 1 lists the coordinates and the U, B, V magnitudes  for all the member stars.

The spectra have been reduced with the dedicated ESO pipelines\footnote{http://www.eso.org/sci//software/pipelines/}, 
including bias-subtraction, flat-fielding, wavelength calibration with a standard Th-Ar lamp, 
spectral extraction and (only for the UVES spectra) order merging. 
The contribution of the sky has been subtracted from each spectrum by using a median sky spectrum 
obtained combining $\sim$15-20 (for GIRAFFE) and 2 (for UVES) spectra of sky regions secured within each exposure.

\section{Analysis}

\begin{figure*}[h]
\plotone{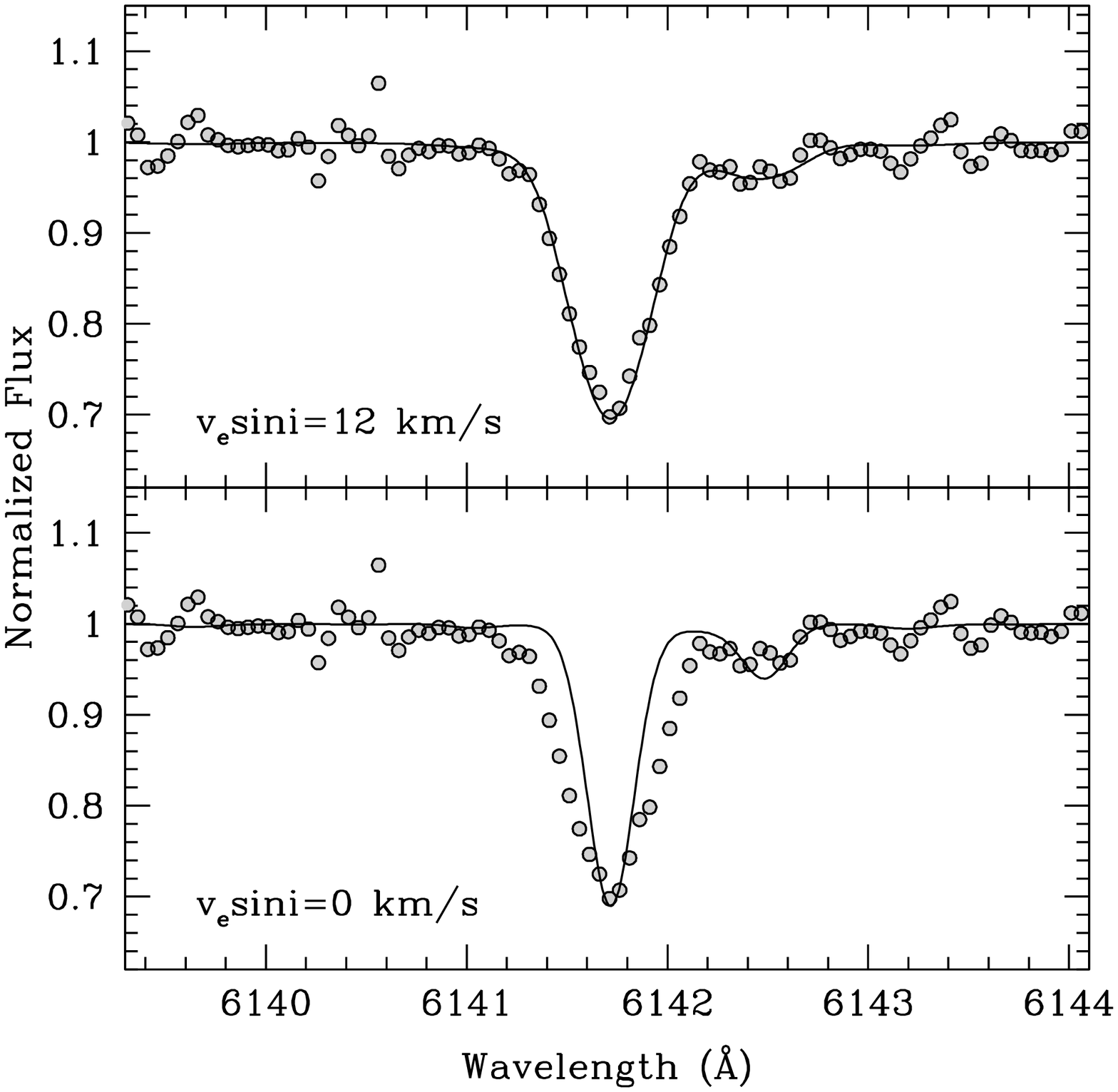}
\caption{Spectral region around the Ba~II line at 6141 \AA\ of the RHB star \#604095 
superimposed with the best-fit synthetic spectra calculating with $v_{e}sini$=~12 km/s 
(upper panel) and vsini=0 km/s (lower panel).}
\label{rotex}
\end{figure*}

The chemical abundances of Fe and Na have been derived with the package {\tt GALA} \citep{m13g}
by matching measured and theoretical equivalent widths (EWs) of a set of unblended lines.
EWs and radial velocities (RVs) have been measured with the code {\tt DAOSPEC} \citep{stetson},  
automatically launched by using the package {\tt 4DAO} \citep{m13_4dao}, that allows a visual inspection of 
all the performed Gaussian fits. 
We started by adopting as input value the nominal full width at half-maximum (FWHM) of the gratings, leaving {\tt DAOSPEC} 
free to re-adjust the FWHM in order to minimize the median value of the residuals distribution 
\citep[see][]{stetson}. 
For some RHB stars the final FWHM is larger than those found for the RGB stars, 
indicating the presence of a rotational broadening in addition to the instrumental profile. 
The projected rotational velocities ($v_{e}sini$) have been measured by comparing the line profiles 
of the RHB stars with grids of suitable synthetic spectra 
calculated with the code {\tt SYNTHE} \citep{sbordone} and convolved with the nominal spectral resolution 
of the used gratings\footnote{These values are 
confirmed also by the measure of the FWHM of bright unsaturated lines in the reference 
Th-Ar calibration lamp, following the method described by \citet{behr00}.}
and with a rotational profile, varying the value of $v_{e}sini$. All the observed RHB stars 
have  $v_{e}sini$ between 0 and 12 km/s. The observed 
distribution of rotational velocities is compatible with those measured for RHB stars in other GCs 
\citep{gratton11,gratton13}. On the other hand, all the RGB stars are compatible 
with zero rotation.
Fig.~\ref{rotex} shows, as example, the spectral region around the Ba~II line at 6141 \AA\ 
for the RHB star \#604095, with superimposed two  synthetic spectra calculated with different rotation 
velocities. In the upper panel the observed spectrum is compared with a synthetic spectrum
calculated with the best-fit rotational velocity ($v_{e}sini$=~12 km/s), while the lower panel 
shows the comparison with a synthetic spectrum without rotation ($v_{e}sini$=0 km/s), not able 
to properly reproduce the line broadening.

First, we measured RV in each individual exposure.
The spectra of each exposure have been corrected for their own heliocentric RV, then 
those corresponding to the same target have been co-added together. Finally the averaged spectra have been analyzed 
to obtain the EWs to be used for the chemical analysis.

Oscillator strengths for the used Fe~I lines are from the critical compilation by \citet{fuhr06}, 
while those for the two Na doublets are from the NIST database 
\footnote{http://physics.nist.gov/PhysRefData/ASD/lines\_form.html}.
Na abundances have been corrected for non-local thermodynamic equilibrium effects adopting the grid of corrections 
calculated by \citet{gratton99}. Model atmospheres have been calculated with the last version 
of the ATLAS9 code\footnote{http://wwwuser.oats.inaf.it/castelli/sources/atlas9codes.html}. 
Solar reference abundances are from \citet{gs98}.

Guess values for the effective temperatures ($T_{\rm eff}$) and surface gravities (log~g) have been 
derived from the photometry (PaperI). In particular, $T_{\rm eff}$ have been estimated by means of the 
$(B-V)_{0}-T_{\rm eff}$ relation by \citet{alonso99}, adopting the extinction law by \citet{cardelli89} 
and a color excess E(B-V)=~0.09 mag \citep{harris}. 
Gravities have been computed through the Stefan-Boltzmann equation, adopting 
the photometric $T_{\rm eff}$, the bolometric corrections by \citet{alonso99}, 
a distance modulus $(m-M)_V=14.68$ mag \citep{harris} and an evolutive mass 
of 0.75$M_{\odot}$, according to a 12 Gyr BaSTI isochrone with Z=0.004 and $\alpha$-enhanced 
chemical mixture \citep{pietr06}. 

Only for the UVES targets $T_{\rm eff}$ can be derived spectroscopically, thanks to the large 
number of available iron lines distributed over a large range of excitation potentials, while this 
approach is not possible for the GIRAFFE targets because of the low number of Fe~I lines with low excitation potential.
For the UVES targets $T_{\rm eff}$ have been derived by imposing the excitation equilibrium, 
i.e. the best $T_{\rm eff}$ is that which erases all trends between Fe~I abundances and excitation potential. 
The photometric $T_{\rm eff}$ of the UVES targets are slightly higher than those derived spectroscopically 
and they need to be lowered by $\sim$80 K to match the spectroscopic ones. 
This offset has been applied to all the targets, including those belonging to the GIRAFFE sample, 
in order to use a self-consistent $T_{\rm eff}$ scale. 
Final values of log~g have been computed with the new $T_{\rm eff}$.

Microturbulent velocities ($v_{\rm turb}$) for RGB stars have been derived spectroscopically 
by erasing all trends between abundances from Fe~I lines and the measured EWs.
For the RHB targets, the small number of available Fe~I lines (less than 25) prevents a robust 
determination of $v_{\rm turb}$. 
According to the analyses of RHB stars in GCs performed by \citet{gratton11,gratton12,gratton13}, 
a unique value of $v_{\rm turb}$ can be used for all the RHB stars of a given cluster. Thus, we assumed for all the 
RHB stars of the sample $v_{\rm turb}$=~1.5 km/s, since this value provides an average [Fe/H] abundance 
for the RHB stars that matches well that obtained for RGB stars.

The abundance uncertainties  have been computed by adding in quadrature two main sources: 
{(i)}~uncertainties arising from the EW measurements, that are estimated 
as the dispersion of the mean normalized to the root mean square of the number 
of used lines. They are of the order of $\sim$0.01-0.03 dex for Fe and smaller than $\sim$0.05-0.06 dex 
for Na; 
{(ii)}~uncertainties arising from the atmospheric parameters. 
Errors in $T_{\rm eff}$ have been estimated taking into account for the photometric uncertainties 
and the uncertainty in the color excess, and they are of about 80-100 K.  Those in 
log~g have been estimated by taking into account the photometric errors and the 
uncertainties in $T_{\rm eff}$, distance 
modulus and mass; they are of the order of 0.05 dex.
A conservative uncertainty of 0.2 km/s has been adopted for the microturbulent velocities.

\section{Fe and Na abundance in NGC~6362}

\begin{figure*}[h]
\plotone{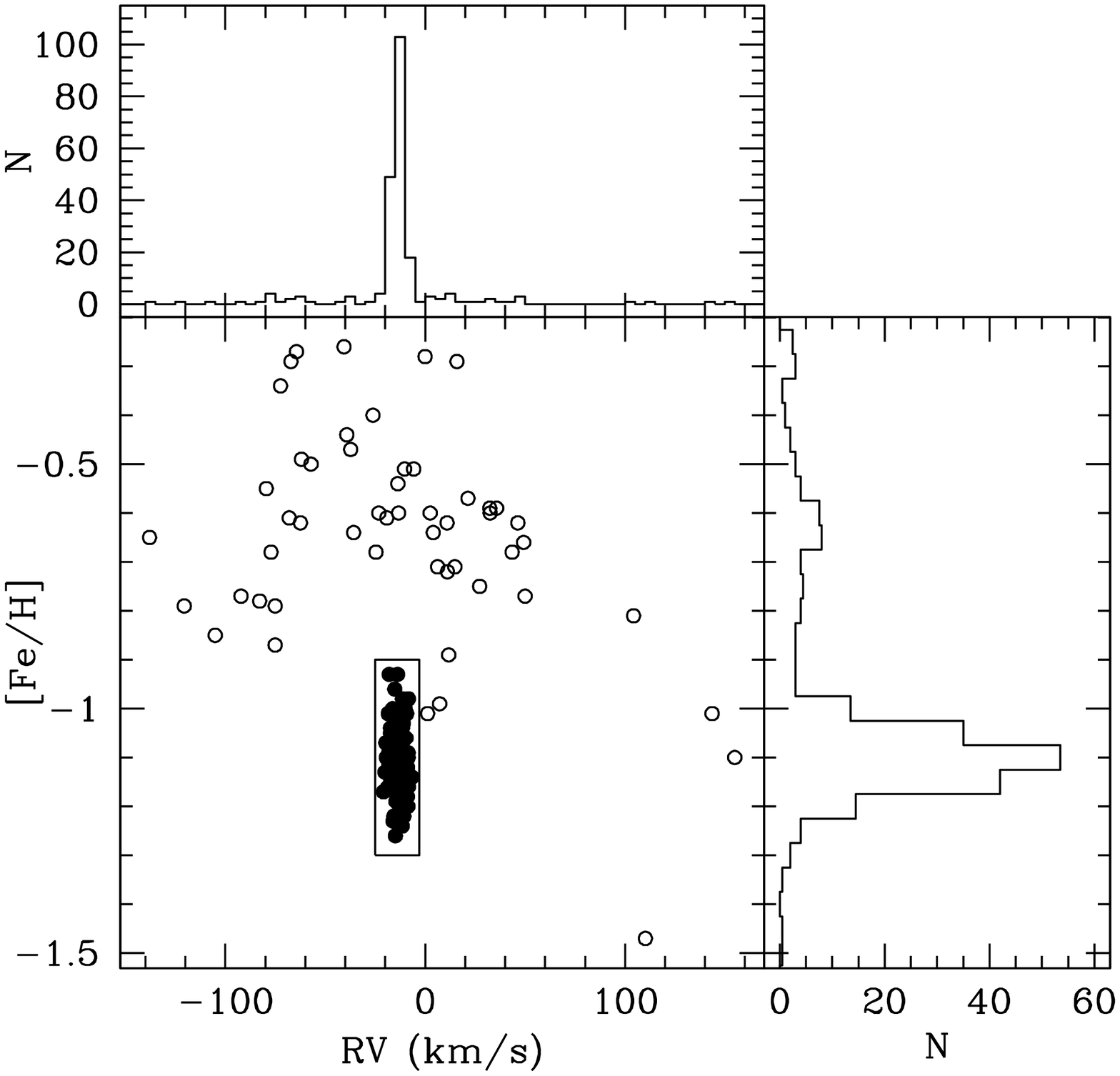}
\caption{Position of the spectroscopic targets in the RV--[Fe/H] plane (same symbols of Fig.~\ref{cmd1}).
The rectangle is the box used to select the bona-fide cluster members. 
The histograms of [Fe/H] and RV distributions are also plotted.}
\label{rvfes}
\end{figure*}

Within the entire sample of observed stars, we identified a total of 160 bona fide cluster members 
(105 RGB and 55 RHB stars), according to their radial velocity and metallicity. 
Fig.~\ref{rvfes} shows the distribution of the observed targets in the RV--[Fe/H] plane, together 
with the histograms of the RV and [Fe/H] distributions. 
Stars belonging to NGC~6362 are easily identified in this plane, since they cluster around 
RV$\sim$--15 km/s and [Fe/H]$\sim$--1.1 dex and they are easily distinguishable from field stars 
because the latter have metallicities peaked at [Fe/H]$\sim$--0.6 dex. 
We selected as cluster members the stars in the range --25$<$RV$<$-3 km/s and  --1.3$<$[Fe/H]$<$-0.9 dex 
(filled circles in Fig.~\ref{rvfes}), excluding, among them, 3 stars with significant RV variations among 
the individual exposures, thus likely being binary systems.
The position in the CMD of the member stars is shown in Fig.~\ref{cmd0} as filled circles, 
while the discarded stars are marked as empty circles.
Table 1 lists the derived [Fe/H] and [Na/Fe] abundance ratios and the corresponding uncertainties 
for each cluster member.

Mean abundances, intrinsic scatters and their uncertainties have been calculated with 
the Maximum Likelihood (ML) algorithm described in \citet{m12}. 
The average iron abundance is [Fe/H]=--1.09$\pm$0.01 dex with an observed scatter of 0.06 dex and an
intrinsic scatter of 0.00$\pm$0.01 dex, 
demonstrating the high degree of iron abundance homogeneity of NGC~6362.
{\sl This is the first determination of the iron content of NGC~6362 based on high-resolution 
spectra}.
Note that the only previous measure of its metallicity has been provided by \citet{geisler97} by using 
Washington photometry, finding [Fe/H]=--0.74$\pm$0.05 dex.

\begin{figure*}[h]
\plotone{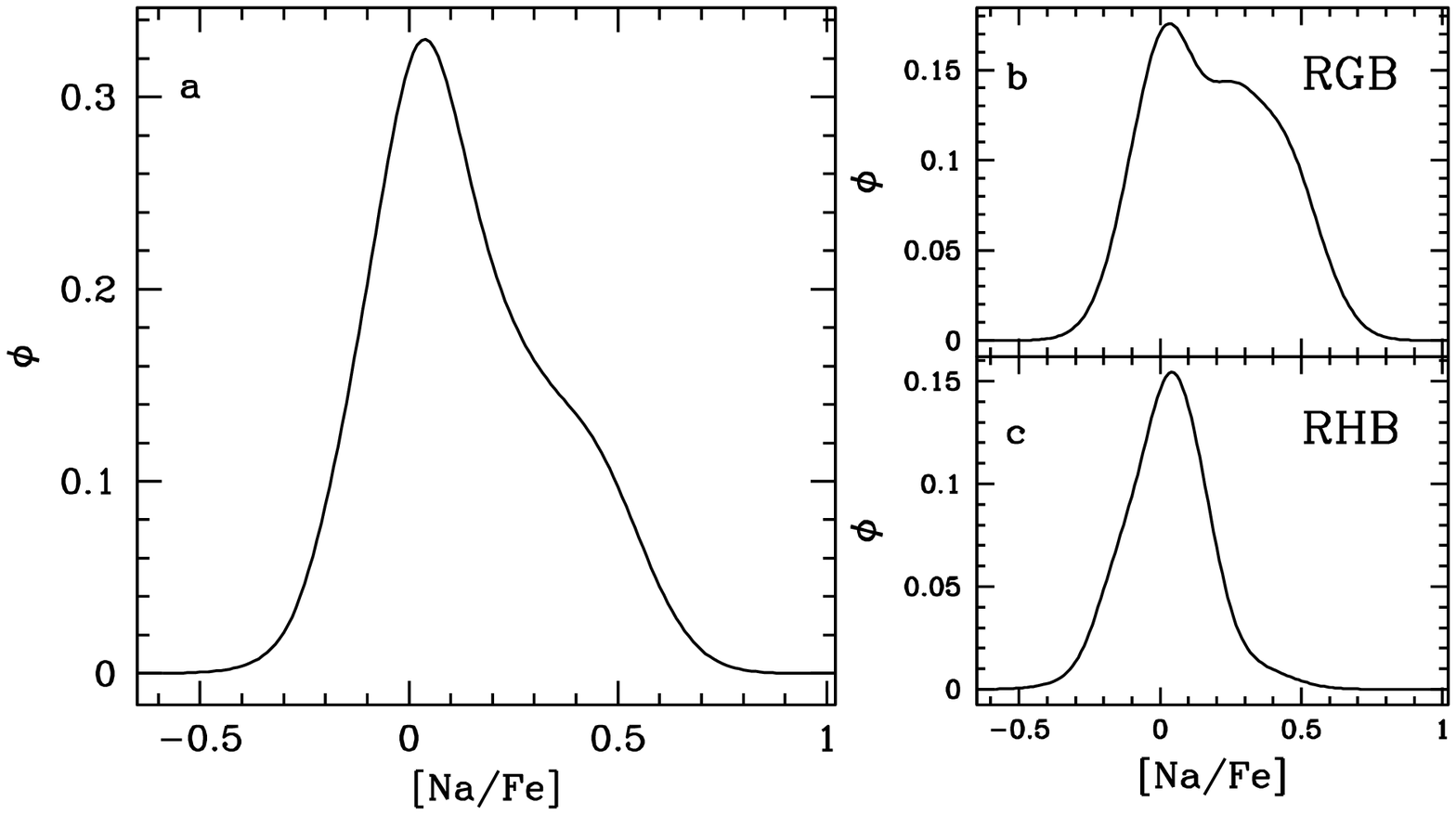}
\caption{Generalized histograms of the [Na/Fe] abundances for the entire sample 
 of member stars (panel a), for the only RGB (panel b) and the only RHB 
 (panel c).}
\label{sodium1}
\end{figure*}

The [Na/Fe] distribution of the sample of member stars is shown in panel {\sl a)} of Fig.~\ref{sodium1} 
as a generalized histogram. The ML algorithm provides an average value of +0.13$\pm$0.01 dex
with an intrinsic dispersion of 0.16$\pm$0.01 dex, indicating a significant star-to-star scatter. 
The [Na/Fe] distributions for the samples of RGB and RHB stars considered separately are shown in 
panels {\sl b)} and {\sl c)} of Fig.~\ref{sodium1}, respectively. The two distributions appear quite different 
from each other: that of RGB stars is about 0.8 dex wide, with a main peak at [Na/Fe]$\sim$0.0 dex, a 
secondary peak close to [Na/Fe]$\sim$+0.3 dex  
and an extended tail up to [Na/Fe]$\sim$+0.6 dex. On the other hand, the distribution of RHB stars is 
narrower than that of RGB stars, with a Gaussian shape and a peak at [Na/Fe]$\sim$0.0 dex, 
nicely matching the main peak of the RGB distribution.

\begin{figure*}
\plotone{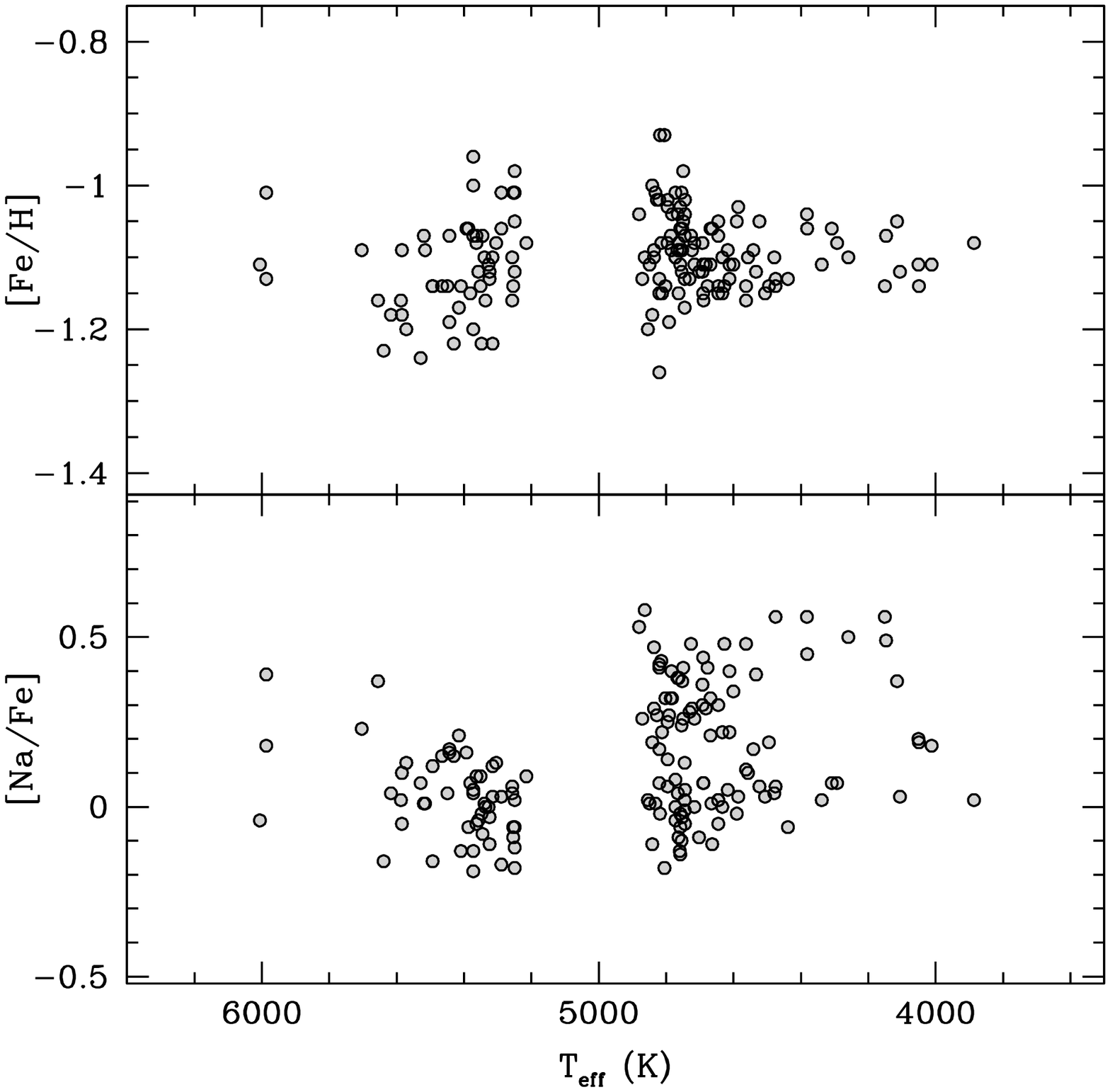}
\caption{Behavior of [Fe/H] (upper panel) and [Na/Fe] (lower panel) as a function of 
$T_{eff}$ for the member stars of NGC~6362.}
\label{tt2}
\end{figure*}

Fig.~\ref{tt2} shows the behavior of [Fe/H] (upper panel) and [Na/Fe] (lower panel) as a function 
of $T_{eff}$ for all the member stars of the cluster: no significant trend between abundance ratios 
and $T_{eff}$ is found.

\section{Na abundance in NGC~6362 and multiple populations}

\subsection{RGB stars}

The bimodality of the [Na/Fe] distribution of the RGB stars has been
statistically analyzed by means of the Gaussian mixture modeling 
algorithm described by \citet{muratov10}. We found that the
hypothesis of unimodal distribution can be rejected with a probability
$>99.9\%$: the observed distribution can be reproduced with two
Gaussian components, one peaked at [Na/Fe]$=+0.00$ dex and with
$\sigma= 0.07$ dex, the other peaked at [Na/Fe]$=+0.33$ dex and with
$\sigma= 0.13$ dex.

Usually, the Na abundance is used to discriminate between first
generation (FG) and second generation (SG) stars, being Na-poor and
Na-rich, respectively.  According to the [Na/Fe] distribution shown in the panel {\sl b)} of
Fig.~\ref{cmd1}, we therefore adopted [Na/Fe]$=+0.15$ dex as the
boundary to separate FG and SG populations in NGC 6362 and, following
the approach already adopted in other works about multiple populations
of GCs \citep[see e.g.][]{marino08}, we investigated the connection
between the two Na-selected samples and the two RGBs observed in the
CMD when the U filter is used. Such a RGB splitting is thought to be
driven by variations of strength of the CN and NH molecular bands that
dominate the $3000-4000 \rm\mathring{A}$ spectral region sampled by
the U filter \citep{sbordone11}, and should therefore be directly
ascribable to multiple populations.

As clearly shown in panel {\sl a)} of Fig. 4, and as expected, the Na-poor
stars reside on the bluest RGB, while the Na-rich stars populate the
reddest RGB. The same result is found if we consider the ($U, U-B$)
plane.  We also found that the fractions of (Na-selected) FG and SG
stars in NGC 6362 are basically the same ($\sim 47\%$ and $\sim 53\%$,
respectively), corresponding to a ratio $N_{\rm SG}/N_{\rm FG}
=1.10\pm 0.21$.  This value agrees well with the second-to-first
generation ratio estimated photometrically in PaperI ($\sim 1.2$)
according to the star number counts along the two RGBs\footnote{ 
We checked that the use of different prescriptions for the NLTE corrections 
for Na abundances does not change these results. [Na/Fe] abundance 
ratios have been re-calculated adopting the NLTE corrections by \citet{lind11}
that provide abundances lower than about 0.15-0.2 dex, while 
the overall [Na/Fe] distribution, the $N_{\rm SG}/N_{\rm FG}$ ratio and the 
photometric distribution of Na-rich and Na-poor stars remain the same.}.

\subsection{HB stars}
\begin{figure*}
\plotone{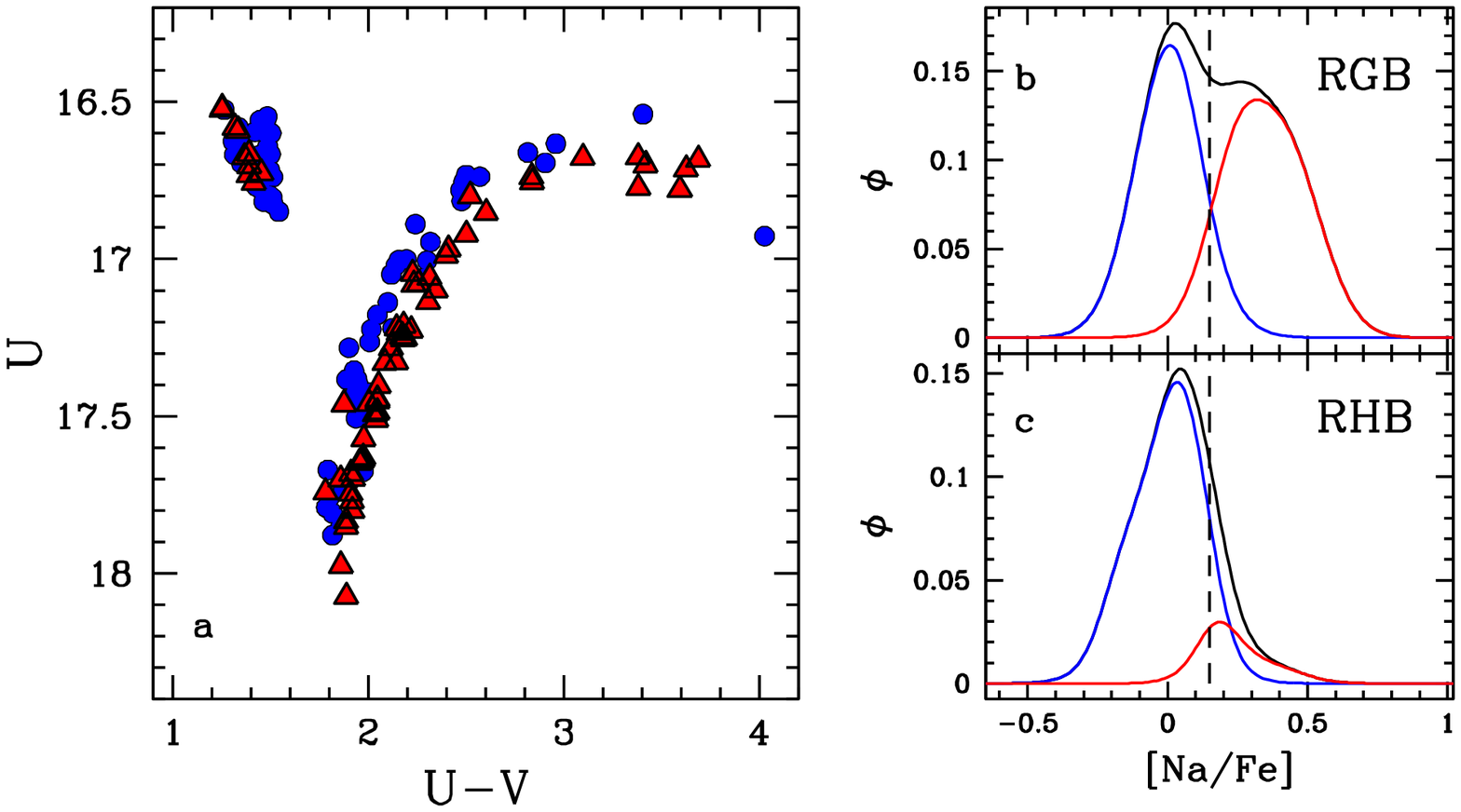}
\caption{Panel {\sl a)}: (U, U-V) CMD of the spectroscopic targets (only member stars). 
The stars are divided according to the [Na/Fe] abundance ratio, assuming [Na/Fe]=~0.15 dex 
as boundary to distinguish between FG and SG stars. 
Red triangles are Na-rich stars, blue circles the Na-poor stars. 
The panel {\sl b)} and {\sl c)} shows the [Na/Fe] distribution of the RGB and RHB stars, respectively, with 
the generalized histograms of the Na-poor and Na-rich stars (blue and red, respectively).}
\label{cmd1}
\end{figure*}

As appreciable in panels {\sl c)} of Fig.~\ref{sodium1} and ~\ref{cmd1}, the [Na/Fe] distribution 
of the RHB stars appears to be quite different from that of RGB stars:
about 82\% of the RHB stars are Na-poor ([Na/Fe]$<$0.15 dex), 
leading to a second-to-first generation ratio  $\frac{N_{\rm SG}}{N_{\rm FG}}$=0.22$\pm$0.08. 
Hence, most of the stars populating the RHB of NGC~6362 belong to the first generation.
This finding confirms  previous analyses of HB stars in other GCs 
\citep[see e.g.][]{marino11,gratton11,gratton12,gratton13}, where the 
red part of the HB is preferentially populated by FG stars.

The natural explanation is that most of the RHB stars that we observe are the progeny 
of stars previously located along the bluest RGB detected in the CMDs including the U filter. 
In this view, the Na-rich RGB stars 
populating the bluest RGB of Fig.~\ref{cmd1} (and accounting for $\sim$50\% of the 
total RGB stars) will populate the blue part of the HB of the cluster. 
Note that  
HB stars bluest than the instability strip have not been observed in this program.

Both the number counts of the two RGBs (Paper I) and the ratio between Na-poor and Na-rich RGB stars 
point out a similar population of FG and SG stars, while the RHB stars are mainly FG stars. 
We calculated the number of Na-rich stars expected to populate the blue HB in order 
to obtain a second-to-first generation ratio of 1.10 along the entire HB. 
The stars have been selected from the WFI catalog within 850'' from the 
cluster center (corresponding to the FLAMES field of view) and adopting (B-V)=~0.4 mag 
as boundary between red and blue HB stars. A total of 130 and 80 red and blue HB, respectively, 
have been counted. The RHB is located in a 
region of the CMD contaminated by main sequence field stars (see Fig.~\ref{cmd0}), at variance 
with the blue portion of the HB. For this reason the number counts of the RHB 
have been corrected for the field star contamination considering a percentage of 
field stars of 19\% (according to the fraction of field stars detected among the RHB spectroscopic targets). 
Finally, we expect to find 79 (out 80) Na-rich blue HB stars. 
Hence, in order to have the same second-to-first generation ratio that we observe 
among the RGB stars, the blue HB should be populated only by Na-rich stars, 
in agreement with the analysis of blue HB stars in GCs performed by \citet{gratton11,gratton12,gratton13}.

\section{Comparison with M4}

Among all the other Galactic GCs with secure evidence of multiple stellar populations, 
M4 is the least massive one with a metallicity comparable to that of NGC~6362 
\citep{harris,marino08,carretta09,villanova11}. 
Hence, a comparison between these two clusters could bring particularly 
interesting information, and, by following the same approach described in \citet{m13_5694},
we therefore adopted the same technique described above to perform the chemical analysis 
of a sample of stars in M4. 
In this way, the derived [Fe/H] and [Na/Fe] 
abundance ratios for the two GCs are from any possible systematics arising from the 
adopted model atmospheres, linelists, method to measure EWs, code to derive the abundances, solar reference 
abundances and NLTE corrections. This kind of comparative analysis is much more robust than a simple 
comparison with literature abundances.

We retrieved, from the ESO Archive\footnote{http://archive.eso.org/cms/eso-data.html}, 
UVES and GIRAFFE spectra of 93 stars along the RGB of M4, acquired 
with the same setups used for NGC~6362 and with comparable 
spectral quality (Prop ID: 072.D-0570, PI: Carretta).
The atmospheric parameters have been computed following the same procedure described 
above. $T_{\rm eff}$ have been derived from B and V photometry \citep[taken from ][]{carretta09}, 
adopting E(B-V)=0.35 mag \citep{harris} and applying suitable corrections for differential reddening 
computed according to the procedure described in \citet{massari12}. Surface gravities have been 
computed assuming a distance modulus of $(m-M)_V$=~12.82 mag \citep{harris}. 
We restrict the analysis only to the stars considered as cluster members by \citet{carretta09}.
The average [Fe/H] of the sample is [Fe/H]=--1.14$\pm$0.01 dex with an observed scatter 
$\sigma_{obs}$=0.06 dex and an intrinsic scatter $\sigma_{int}$=0.00$\pm$0.02 dex, 
in nice agreement with the previous determinations \citep[see e.g.][]{marino08,carretta09,mucciarelli11} 
and confirming that M4 and NGC~6362 have very similar metallicities.

\begin{figure*}
\plotone{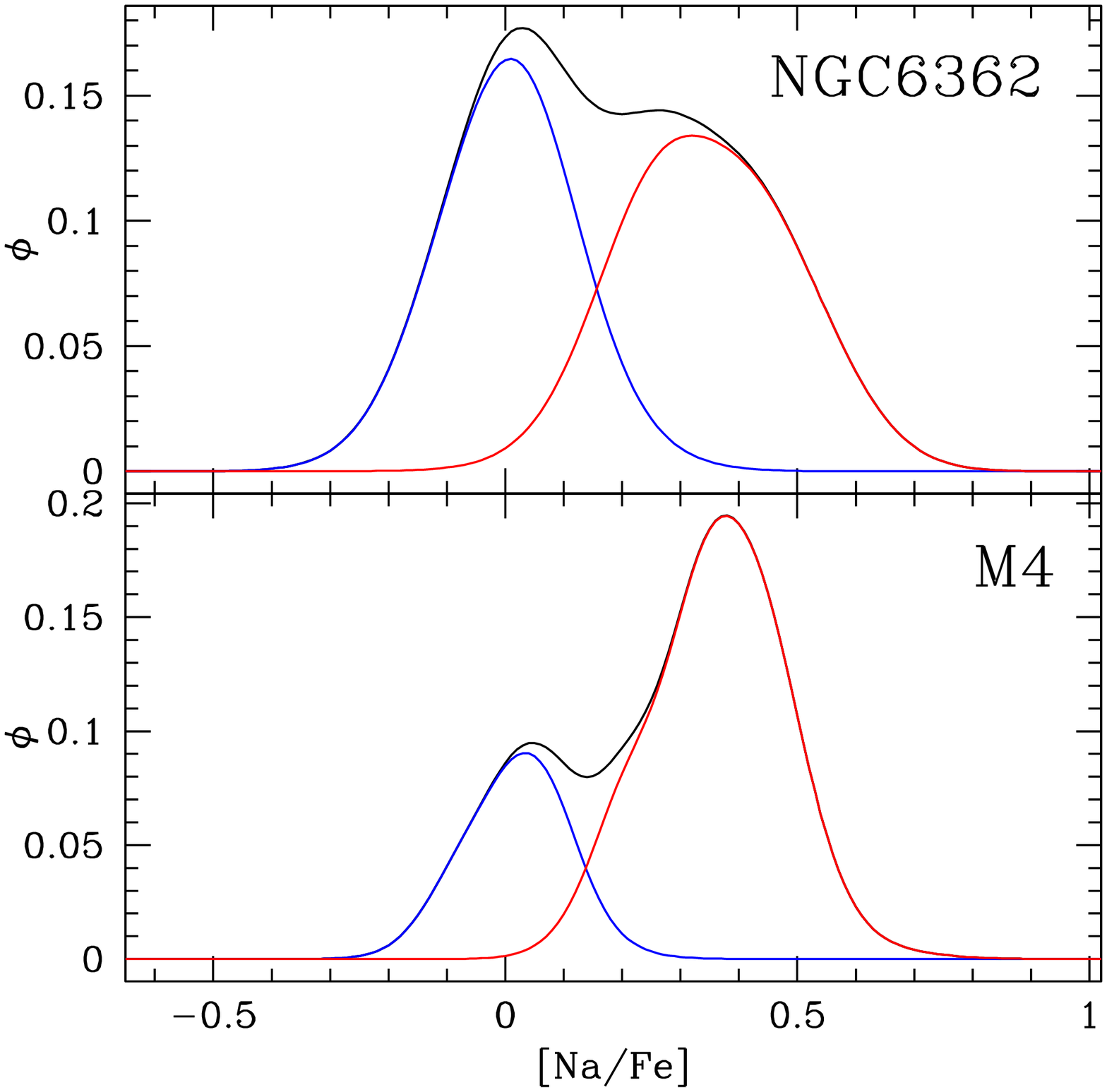}
\caption{[Na/Fe] distribution in the RGB stars of NGC~6362 (upper panel) and M4 (lower panel). 
In both panels, blue and red histograms are the distributions of Na-poor and Na-rich stars, respectively, 
assuming [Na/Fe]=0.2 dex as boundary.}
\label{comp}
\end{figure*}

The distribution of [Na/Fe] of M4 stars spans about 0.7 dex, with 
an average value of +0.27$\pm$0.01 and an intrinsic spread of 0.16$\pm$0.01.
The comparison between the [Na/Fe] distributions of NGC~6362 and M4 is shown in Fig.~\ref{comp}. 
The two distributions are similar in terms of the covered abundance range and 
the position of the two main peaks at [Na/Fe]$\sim$0.0 and +0.3 dex.
However we note that they show a remarkably different  
number ratio of the two populations, with the Na-rich population dominating 
in the case of M4. In fact, 
assuming the same boundary adopted for NGC~6362, the ratio between Na-rich and Na-poor stars in M4 is 
$\frac{N_{\rm SG}}{N_{\rm FG}}$=~2.76$\pm$0.70, corresponding 
to a fraction of FG stars of $\sim$27\%. This is in agreement with 
what found by \citet{marino08} and \citet{carretta09} for M4, and to that 
usually found in all GCs 
(\citet{bastian15b} estimated an average ratio between Na-rich and Na-poor stars of 0.68$\pm$0.07\%).

\section{Summary and conclusions}

The main results obtained from the spectroscopic analysis presented in this paper are:\\
{\sl (i)}~the iron content of NGC~6362 is [Fe/H]=--1.09$\pm$0.01, without evidence of 
intrinsic spread. This is the first determination of its iron abundance based on high-resolution spectra;\\
{\sl (ii)}~the [Na/Fe] distribution of the cluster shows a clearly bimodal distribution, revealing 
the presence of two well distinct stellar populations.
We find a clearcut correspondence between the Na abundances and the 
two RGBs detected in the CMDs including the U or the F336W HST filters (PaperI), 
with the Na-poor stars populating the bluest RGB and the Na-rich stars the reddest RGB, as theoretically
expected;\\
{\sl (iii)}~among the RGB stars, the two populations, selected according to their [Na/Fe] abundance, 
are equally populated, at variance with the other GCs where the stellar content is dominated by the Na-rich population.

The results of this paper confirm that NGC~6362 is the least massive cluster where MPs have 
been observed, as discussed in PaperI on a pure photometric basis. 
The other two GCs with comparable masses and evidence of chemical 
anomalies,
namely NGC~288 and M4, are 2 and 3 times more massive than NGC~6362, respectively, 
when masses are estimated in a homogeneous way (see PaperI), by using the best-fit 
King model reproducing their density profiles and the velocity dispersion data 
by \citet{pryor93}.
On the other hand, old open clusters slightly 
less massive (by a factor of 3-5) than NGC~6362 do not exhibit intrinsic scatters in the light elements, 
see the case of Berkeley 39 \citep{bragaglia12} and NGC~6791, though this latter
case is still matter of debate since different groups find opposite results 
\citep{geisler12,bragaglia14,cuhna15}. Whatever the true nature of the polluters is, 
this finding about NGC~6362 can provide crucial constraints  
to the minimum cluster mass necessary to retain the low-energy ejecta and to undergo 
self-enrichment processes.

The differential comparison with M~4 clearly reveals that these two clusters share the same chemical patterns, 
with broad and bimodal [Na/Fe] distributions.
However we observe that in NGC~6362 FG and SG are equally populated at variance with what observed in 
M~4, where SG stars represent the $73\%$ of the total. 
Indeed, the analysis of 19 GCs discussed by \citet{carretta09} 
provides a typical fraction of $\sim$30\% for the first generation and of $\sim$70\% 
for the polluted stars. 
The only GC in the \citet{carretta09} sample with a comparable 
first generation fraction is NGC~2808 (50\%).
NGC~2808 is however very different from NGC~6362 and most GCs, 
in terms of anti-correlation extension \citep{carretta09}, fraction of the extreme generation ([O/Na/$<$--0.9 dex),
horizontal branch morphology \citep{dalex11} and mass 
\citep[NGC~2808 is more massive by a factor of 8, according to][]{mcl05}.

The fact that SG stars typically outnumber FGs is usually explained by invoking a relevant and preferential 
loss of FG stars during the early stages of the GC life. In fact, 
it is expected \citep{dercole11,bekki11} that FG are initially less 
segregated than SG stars. As a consequence, they reach 
the conditions to escape the cluster during the early expansion of the system taking place during 
the first $\sim1 $Gyr of the cluster, which is dominated by stellar evolution and SNe explosions.

In the case of NGC~6362, 
the present-day number ratio between FG and SG stars is representative of the one shown by the cluster 
at the time when FG and SG stars have reached fully spatial mixing, as shown by \citet{vesperini13} and \citet{miholics15}. 
In fact, after mixing is reached, different sub-populations 
are expected to lose stars at similar rates. 
How the present-day ratio is related to the primordial one depends on the fraction 
of mass lost. Both \citet{vesperini13} and \citet{miholics15} find that in order 
to reach a completely spatially mixed configuration, NGC6362 should have lost a 
significant fraction of mass during the long-term dynamical evolution. 
Indeed \citet{miholics15} have estimated that, given its orbit, NGC6362 has been 
tidally filled for most of its life, independently on its initial conditions, 
and therefore it likely lost a large fraction of stars because of tidal stripping. 
If this is the case, NGC6362 is expected to be born with a very small and highly concentrated 
SG sub-population.

We can  conclude that, 
putting together the number of peculiar properties observed in this cluster,  NGC~6362 
results to be an unique case among the Galactic GCs. 
Any theory aimed at explaining the formation and evolution of GCs 
(from initial, more massive clusters to those that we currently observe) must 
account also for the peculiar case of NGC~6362.

\acknowledgements  
We warmly thank the anonymous referee for suggestions that helped improving the paper.
This research is part of the project COSMIC-LAB (http://www.cosmic-lab.eu) funded 
by the European Research Council (under contract ERC-2010-AdG-267675).

\begin{landscape} 
\begin{deluxetable}{lcccccccccccc}
\tablecolumns{13} 
\tablewidth{0pc}  
\tablecaption{}
\tablehead{ 
\colhead{Star} &   RA & Dec  & U & B &  V  &  $T_{\rm eff}$ & logg  & $v_{turb}$ &   [Fe/H] & [Na/Fe]\\
  &   (J2000)  &  (J2000)  &  &  &   &  (K)  &    & (km/s)   & (dex) & (dex)  }
\startdata 
\hline 
201343 &   263.0750664  &   -66.8997455  &  17.220    &  16.048 &  15.074  & 4716  &  2.08    &  1.40	&  -1.11$\pm$0.10   &  0.26$\pm$0.08	\\    
500473 &   262.4560855  &   -67.0569192  &  17.383    &  16.451 &  15.494  & 4745  &  2.27    &  1.60	&  -1.13$\pm$0.10   &  0.02$\pm$0.16	\\	
600257 &   262.8930198  &   -66.9839966  &  17.059    &  15.817 &  14.768  & 4590  &  1.89    &  1.40	&  -1.05$\pm$0.08   & -0.02$\pm$0.11	\\    
600352 &   262.8912788  &   -66.9723982  &  16.799    &  15.387 &  14.278  & 4495  &  1.64    &  1.50	&  -1.14$\pm$0.08   &  0.19$\pm$0.09	\\   
600496 &   262.8879197  &   -67.0502011  &  16.628    &  15.944 &  15.296  & 5587  &  2.55    &  1.50	&  -1.16$\pm$0.10   &  0.02$\pm$0.08	\\   
600530 &   262.8872475  &   -67.1038652  &  17.852    &  16.855 &  15.964  & 4863  &  2.51    &  1.40	&  -1.10$\pm$0.10   &  0.58$\pm$0.10	\\   
600607 &   262.8857838  &   -66.9837179  &  16.733    &  15.354 &  14.232  & 4475  &  1.61    &  1.40	&  -1.13$\pm$0.09   &  0.06$\pm$0.11	\\  
600850 &   262.8802788  &   -67.0446788  &  16.591    &  15.885 &  15.257  & 5655  &  2.56    &  1.50	&  -1.16$\pm$0.12   &  0.37$\pm$0.10	\\  
601019 &   262.8762369  &   -67.0575666  &  17.227    &  16.010 &  15.008  & 4668  &  2.03    &  1.40	&  -1.11$\pm$0.09   &  0.32$\pm$0.10	\\  
601528 &   262.8658176  &   -67.0464912  &  16.667    &  15.922 &  15.166  & 5254  &  2.37    &  1.50	&  -1.14$\pm$0.15   & -0.09$\pm$0.07	\\     
601632 &   262.8636925  &   -66.9832167  &  17.418    &  16.418 &  15.476  & 4772  &  2.27    &  1.30	&  -1.10$\pm$0.10   &  0.00$\pm$0.11	\\    
601756 &   262.8604542  &   -67.0578518  &  16.781    &  15.411 &  14.309  & 4506  &  1.66    &  1.50	&  -1.15$\pm$0.08   &  0.03$\pm$0.09	\\	
602034 &   262.8537861  &   -67.0442901  &  16.525    &  15.793 &  15.261  & 6006  &  2.68    &  1.50	&  -1.11$\pm$0.09   & -0.04$\pm$0.09	\\    
602223 &   262.8486807  &   -67.1455510  &  16.657    &  15.906 &  15.181  & 5345  &  2.41    &  1.50	&  -1.07$\pm$0.12   & -0.08$\pm$0.12	\\   
602471 &   262.8423465  &   -67.0321127  &  16.755    &  16.025 &  15.306  & 5363  &  2.47    &  1.50	&  -1.07$\pm$0.13   & -0.05$\pm$0.08	\\   
602685 &   262.8375124  &   -66.9349993  &  16.815    &  15.429 &  14.338  & 4523  &  1.68    &  1.40	&  -1.05$\pm$0.08   &  0.06$\pm$0.10	\\   
602803 &   262.8337385  &   -67.0625221  &  17.698    &  16.681 &  15.775  & 4836  &  2.42    &  1.40	&  -1.10$\pm$0.10   &  0.29$\pm$0.11	\\  	   
\hline
\enddata 
\tablecomments{$~~~~~$Coordinates, magnitudes, atmospheric parameters, [Fe/H] and [Na/Fe] abundance ratios for the target member stars.
Identification numbers are from PaperI. The entire table is available in the electronic 
version of the journal.}
\end{deluxetable}
\end{landscape}

\end{document}